\documentclass[preprint,12pt]{aastex}

\begin{document}
	\title{SN 2001em: Not so Fast }
	\author{F. K. Schinzel\altaffilmark{1}, G. B. Taylor\altaffilmark{1,2},  C.J. Stockdale\altaffilmark{3}, J. Granot\altaffilmark{4}, and E. Ramirez-Ruiz\altaffilmark{5}}
	\email{schinzel@mpifr.de}

	\altaffiltext{1}{Department of Physics and Astronomy, University of New Mexico, 800 Yale Blvd NE, Albuquerque, NM 87131, USA}
	\altaffiltext{2}{G.B. Taylor is also an Adjunct Astronomer at the National Radio Astronomy Observatory}
	\altaffiltext{3}{Marquette University, Physics Department, P.O. Box 1881, Milwaukee, WI 53214-1881, USA}
	\altaffiltext{4}{Centre for Astrophysics Research, University of Hertfordshire, College Lane, Hatfield, Herts, AL10 9AB, UK}
	\altaffiltext{5}{Department of Astronomy and Astrophysics, University of California, Santa Cruz, CA 95064, USA}

	\received{08/13/2008}
	\revised{09/30/2008}
	\accepted{10/02/2008}
	\cpright{AAS}{2008}

	\begin{abstract}

		SN~2001em is a peculiar supernova, originally classified as Type Ib/c. About two years after the SN it was detected in the radio, showing a rising radio flux with an optically thin spectral slope, and it also displayed a large X-ray luminosity ($\sim 10^{41}\;{\rm erg\;s^{-1}}$). Thus it was suspected to harbor a decelerating (by then, mildly) relativistic jet pointing away from us.  About $3$ years after its discovery the optical spectrum of SN~2001em showed a broad H$\alpha$ line, and it was therefore reclassified as Type IIn. Here we constrain its proper motion and expansion velocity by analyzing four epochs of VLBI observations, extending out to 5.4 years after the SN. The supernova is still unresolved 5.4 years after the explosion. For the proper motion we obtain (23,000$\,\pm\,$30,000)$\;$km$\;$s$^{-1}$ while our 2$\sigma$ upper limit on the expansion velocity is 6000$\;$km$\;$s$^{-1}$. These limits are somewhat tighter than those derived by Bietenholz \& Bartel, and confirm their conclusion that late time emission from SN~2001em, a few years after the explosion, is not driven by a relativistic jet. VLA observations of the radio flux density, at 8.46 GHz, show a decay as $t^{-1.23\pm 0.40}$ starting $\sim 2.7\;$years after the SN. Collectively, the observations suggest interaction of the SN ejecta with a very dense circumstellar medium, though the implied opacity constraints still present a challenge.

	\end{abstract}


\section{Introduction}

		On September 15, 2001, \citet{2001IAUC.7722....1P} reported the discovery of an apparent Supernova (SN) in the galaxy UGC 11794, also mistakenly called NGC~7112 (RA: 21h42m22.9s DEC: 12d29m54s; z=0.019493 J2000.0). Throughout the paper we use H$_0$ = 70~km~s$^{-1}$~Mpc$^{-1}$, and $\Omega_\Lambda = 1-\Omega_M = 0.72$ \citep{2008arXiv0803.0732H}, for which the measured redshift corresponds to a proper distance of $D_p = 83.14\;$Mpc and an angular distance of $D_A = 81.55\;$Mpc. Early spectral line data suggested it was of type I b/c (more likely Ic) \citep{2001IAUC.7737....3F}. It was detected in the radio about $2$ years after the explosion \citep{2004IAUC.8282....2S}, initially displaying a rising flux together with an optically thin spectral slope $F_\nu \propto t^{1.9\pm 0.4} \nu^{-0.36 \pm 0.16}$ ($4.9-15\;$GHz). It also showed a high X-ray luminosity of $\sim 10^{41}\;{\rm erg\;s^{-1}}$ at 0.5$-$8$\;$keV \citep{2004IAUC.8323....2P}. This made it a good candidate for harboring a bipolar relativistic jet pointing away from our line of sight \citep{2004ApJ...609L...9G}, which has by then decelerated to mildly relativistic velocities. Such late time radio emission is expected in some Type Ib/c SNe that may be associated with gamma-ray bursts \citep{Pacz01,GL03} or alternatively produce only mildly relativistic bipolar outflows, and could naturally produce a rising radio flux in the optically thin regime \citep{Granot02}, as the jets interact with the circumstellar medium (CSM) and are decelerated.

		Later observations, about $3$ years after the explosion \citet{2004GCN..2586....1S}, showed broad H$\alpha$ (FWHM 40~\AA\,= 1800~km s$^{-1}$) lines that are not typical of Type Ib/c SN. Observations using Very Long Baseline Interferometry (VLBI) were proposed to resolve the source or detect proper motion if it is powered by decelerating relativistic jets \citep{2004ApJ...609L...9G}. However, despite several attempts with various instruments the source has not been resolved \citep{2005MmSAI..76..570P, 2005ApJ...625L..99B, 2007arXiv0706.3344B, 2005IAUC.8472....4S}. An alternative explanation was suggested by \citet{2006ApJ...641.1051C}, who developed a model in which regular Newtonian ejecta from the supernova explosion collides with a dense, massive circumstellar shell (CS).

		In addition to previously published VLBI observations, we add the most recent High Sensitivity Array (HSA) observation from February 2007 (about 5.4 years after the explosion). In order to provide the most accurate measurements and a conclusive picture for the proper motion, as well as the expansion velocity of SN 2001em, we use our observation in combination with those of \citet{2004AAS...205.7107S} and \citet{2005ApJ...625L..99B,2007arXiv0706.3344B}.

	\section{Observations \& data reduction}

		We observed SN 2001em with the HSA, which consists of NRAO's Very Long Baseline Array (VLBA), the phased Very Large Array (VLA), the Robert C. Byrd Green Bank Telescope (GBT), the Arecibo Radio Telescope (AR), and the Effelsberg Radio Telescope (EB). The observation was conducted on February 5th, 2007 with a total time of 10 hours at 8.4 GHz (bandwidth 32 MHz, total recording bit rate 256 Mbit s$^{-1}$). VLBI data was correlated with the NRAO VLBA processor at Socorro. Analysis was done with NRAO's Astronomical Image Processing System (AIPS) and Caltech's Difmap for modeling. In addition we also used the VLA (in A configuration) observation on February 18th 2007, which observed the continuum of SN 2001em in L, C and K-Band over a total time of about an hour. The three previous VLBA/HSA observations from July 2004 \citep{2004AAS...205.7107S}, November 2004 \citep{2005ApJ...625L..99B} and May 2006 \citep{2007arXiv0706.3344B} were re-reduced in a consistent fashion. A summary of the observations is presented in Table~\ref{tab:observation} including important parameters of these observations.

		The data was corrected for ionospheric delay, Faraday rotation, and parallactic angle as well as for Earth's orientation. We used fringe fitting to calibrate the data for group delay and phase rate. The three datasets from previous VLBI observations have been treated equally, to ensure comparable results. Our VLBI observation of SN 2001em was phase-referenced to JVAS J2145+1115 (1.4$^\circ$ away) and J2139+1423 (2.1$^\circ$ away). The switching intervals were $\sim$180 s for J2145+1115, $\sim$93 s for SN~2001em, and $\sim$39 s for J2139+1423, which was observed in addition about every 5th cycle. We used J2145+1423 to check our phase reference (J2145+1115) to get a measure for the quality of our phase referencing as well as a position fixed to the International Celestial Reference Frame (ICRF). In addition we used J2139+1423 to check astrometry and over results for possible motion within J2145+1115.

	\section{Results}
		
	\subsection{Position and Proper Motion}
		We obtained the results for absolute position, thus for proper motion, by fitting the fully calibrated and phase-referenced ($u$,$v$) plane data, of the observations from 4 epochs, to a circular Gaussian model. This is a good fit for a point source, as we did not see any resolved extended emission. However, the first VLBI observation from July 2004 does not use J2139+1423 as phase reference check as described above, instead it uses J2139+1316 which is 1.0$^\circ$ away from our pointing center. Furthermore only the VLBA was used in this initial observation.

\begin{figure}
 		\plotone{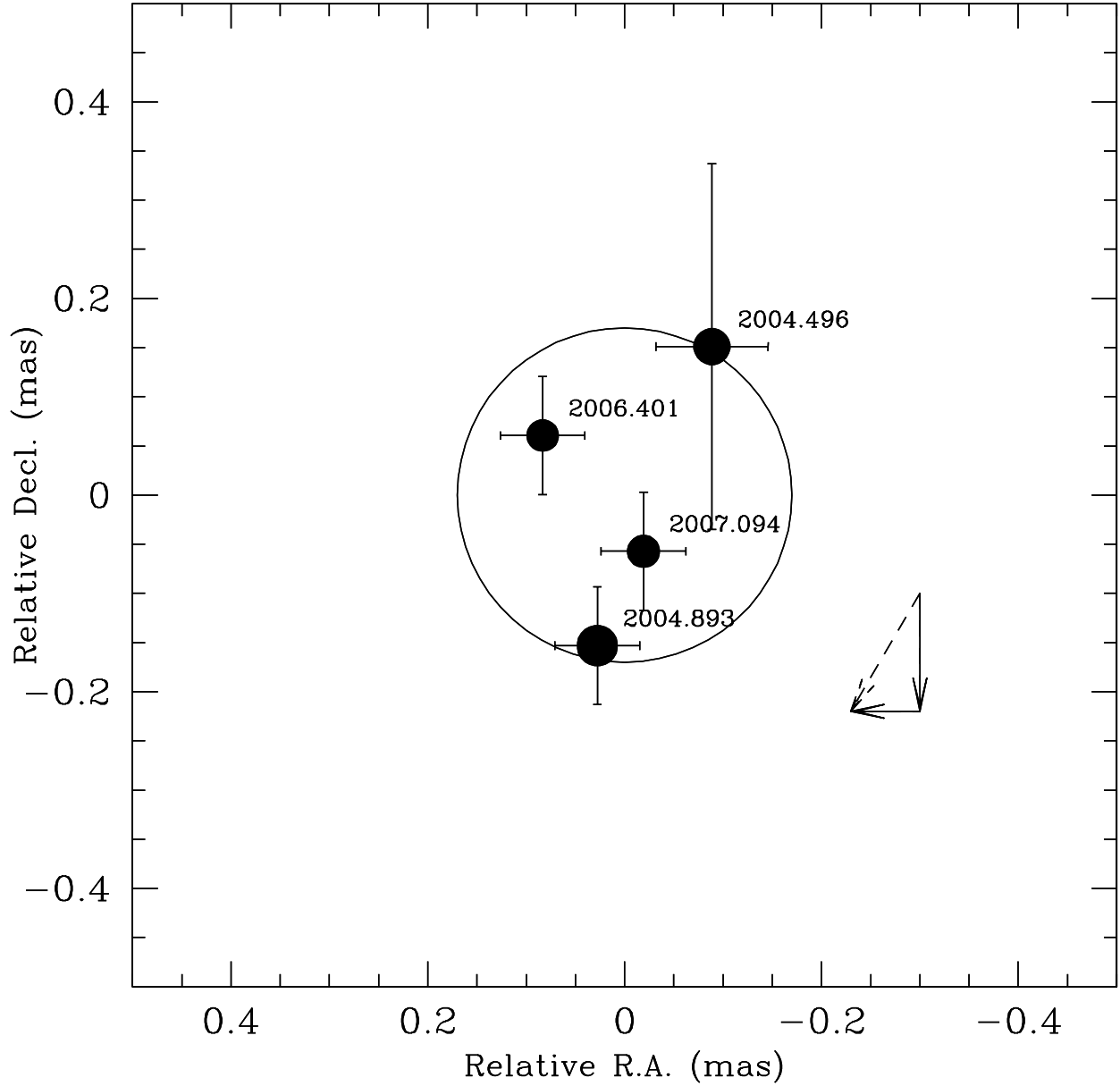}
		\figcaption{\label{fig:sn2001em-pos} Positions of SN 2001em from 2004 to 2007, centered at R.A. 21h42m23.619s Dec. 12d29m50.299s (J2000). Circle centered at the origin has a radius of 0.170 mas. The triangle on the middle right shows the projected velocity vectors in R.A. and Dec. respectively, as well as the total projected velocity for a period of 2.201 years (3 epochs of HSA observations).}
	\end{figure}

		We plot the position including errorbars in Fig.~\ref{fig:sn2001em-pos}. We find a circle with a radius of 0.170~mas that encompasses all the observed positions. This suggests that SN~2001em shows little or no proper motion. 

\begin{figure}
 		\plotone{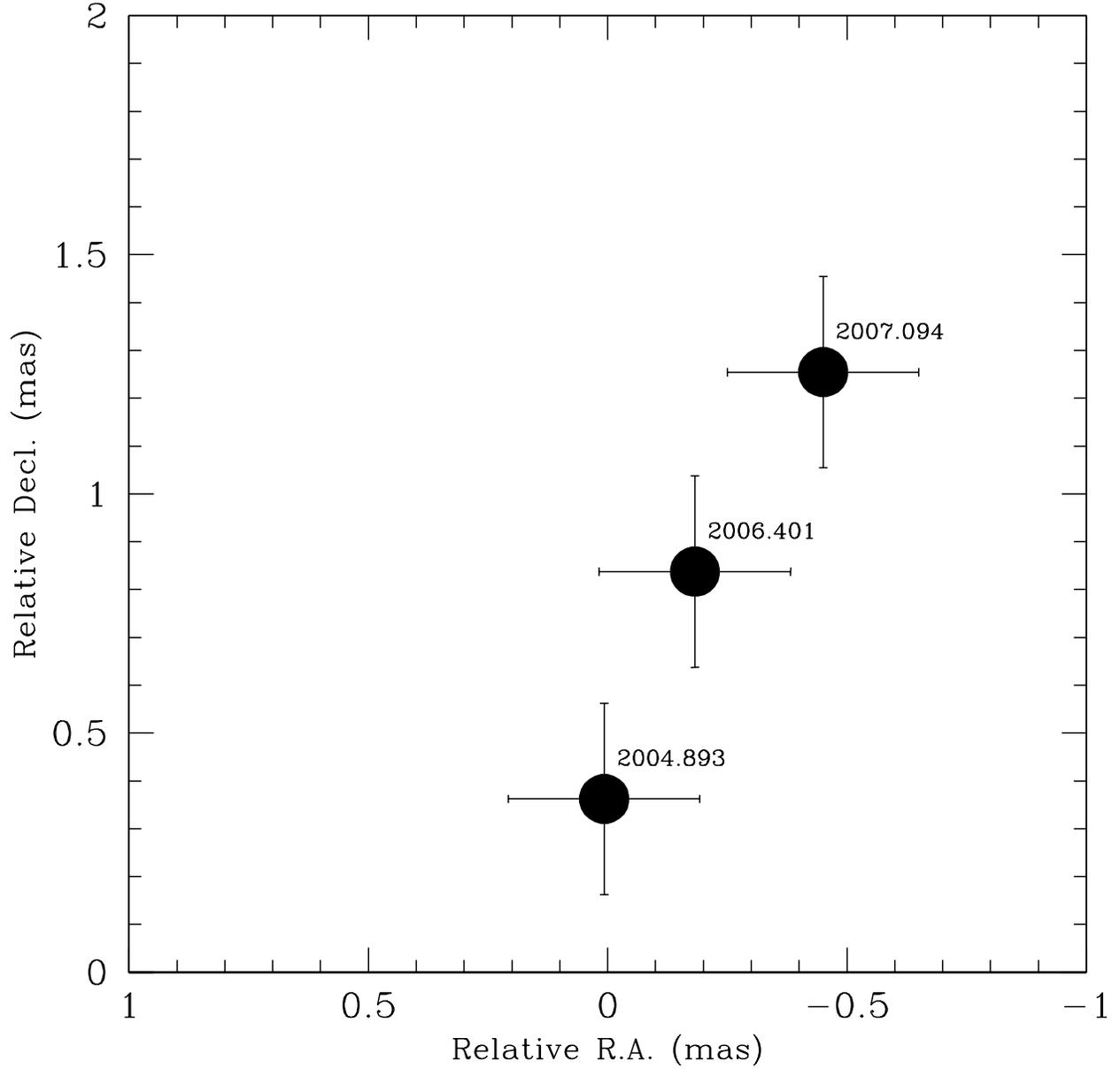}
		\figcaption{\label{fig:j2139+14-pos} Positions of the phase check calibrator, J2139+1423, from 2004 to 2007. 8 epochs of VLBA observations show an apparent relative motion of its components between 0.108 and 0.593 mas year$^{-1}$ \citep{2007AJ....133.2357P}.}
	\end{figure}

\begin{figure}
 		\plotone{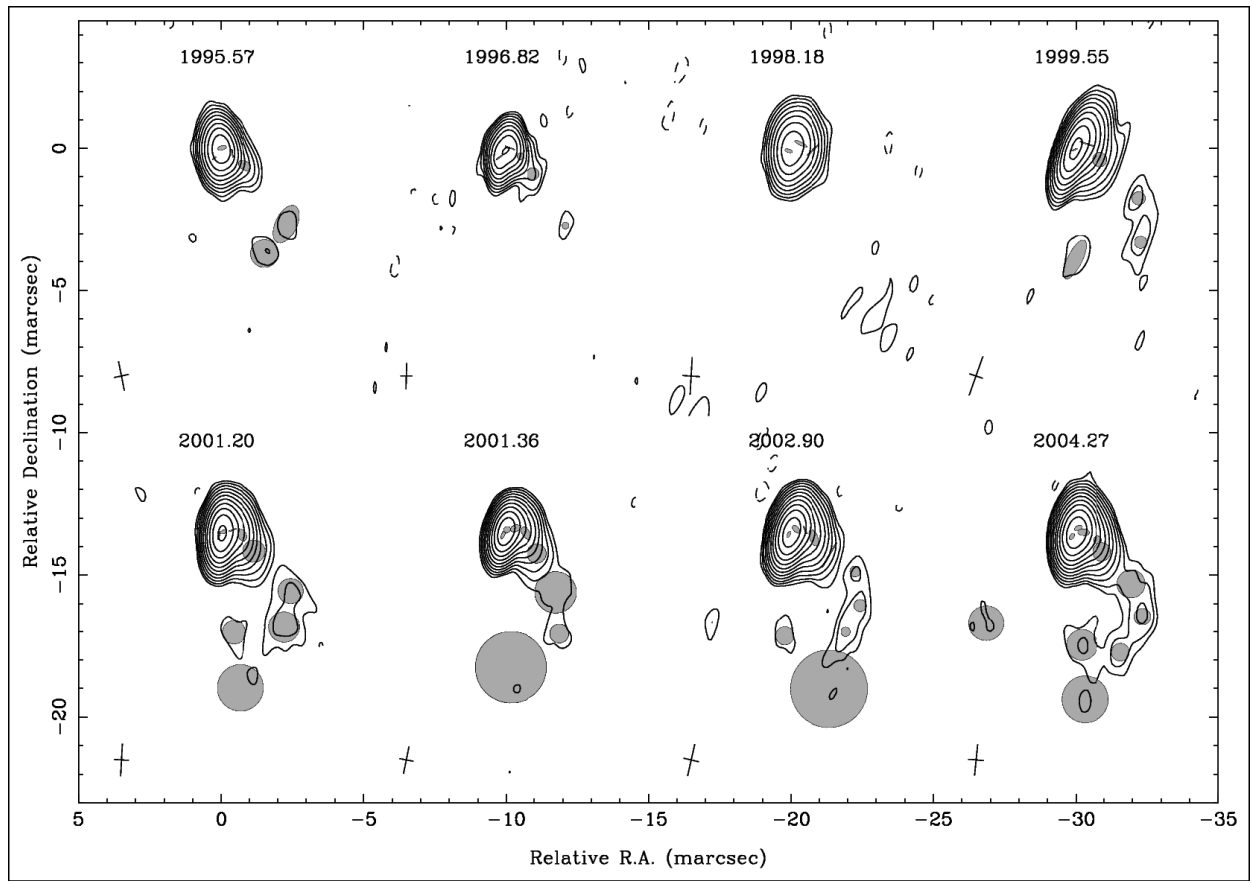}
		\figcaption{\label{fig:2136+141} VLBA images of J2139+1423 (PKS 2136+141) at 15 GHz. The figure includes observations from the VLBA 2 cm Survey, from the MOJAVE Survey and a 15 GHz image from the multi-frequency data set observed in 2001.36 by \cite{2006ApJ...647..172S}. In all images, a uniformly weighted ($u$,$v$)-grid is used. The Gaussian components fitted to the visibility data are shown as ellipses overlaid on each image. The size and orientation of the beam is shown in the lower left corner of each image \citep{2006ApJ...647..172S}. The images show clearly the existence of an extremely curved jet with an inner position angle ranging from -30$^\circ$ to -90$^\circ$.}
	\end{figure}

		To get an estimate for the error of the proper motion, we plotted the positions of J2139+1423 phase referenced to J2145+1115 as shown in Fig.~\ref{fig:j2139+14-pos} and Fig.~\ref{fig:2136+141}, which shows according to our data a proper motion of 0.45 mas year$^{-1}$ in the northwest direction. \cite{2007AJ....133.2357P} give a relative motion of the 4 components between 0.108 and 0.593 mas year$^{-1}$ separated up to 3.5 mas from the core. This can be explained by changes in the extended emission of J2139+1423, which shows an extremely curved jet \citep{2006ApJ...647..172S}, in combination with any proper motion of the phase check calibrator. The intrinsic motion of J2139+1423 has been determined to be $<$ 40 $\mu$as year$^{-1}$ \citep{2003A&A...403..105F}. We also assume our reference source J2145+1115 to be stationary. Using a linear relationship between the stationary phase center and the apparently moving J2139+1423, that are 3.49$^\circ$ apart from each other, we find an error for the proper motion of SN 2001em of $\pm$0.184 mas year$^{-1}$. 
		
\begin{figure}
	 	\plottwo{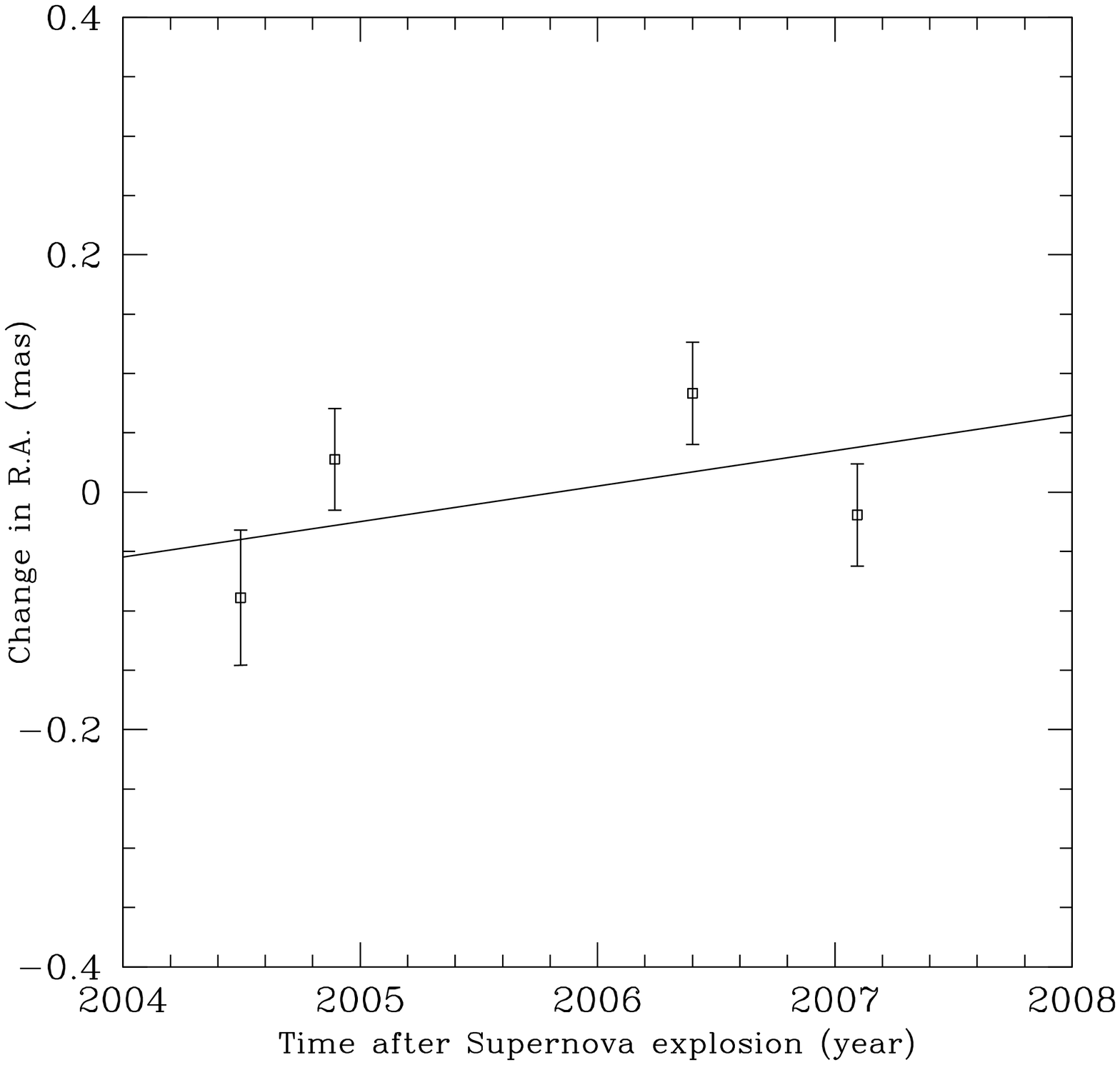}{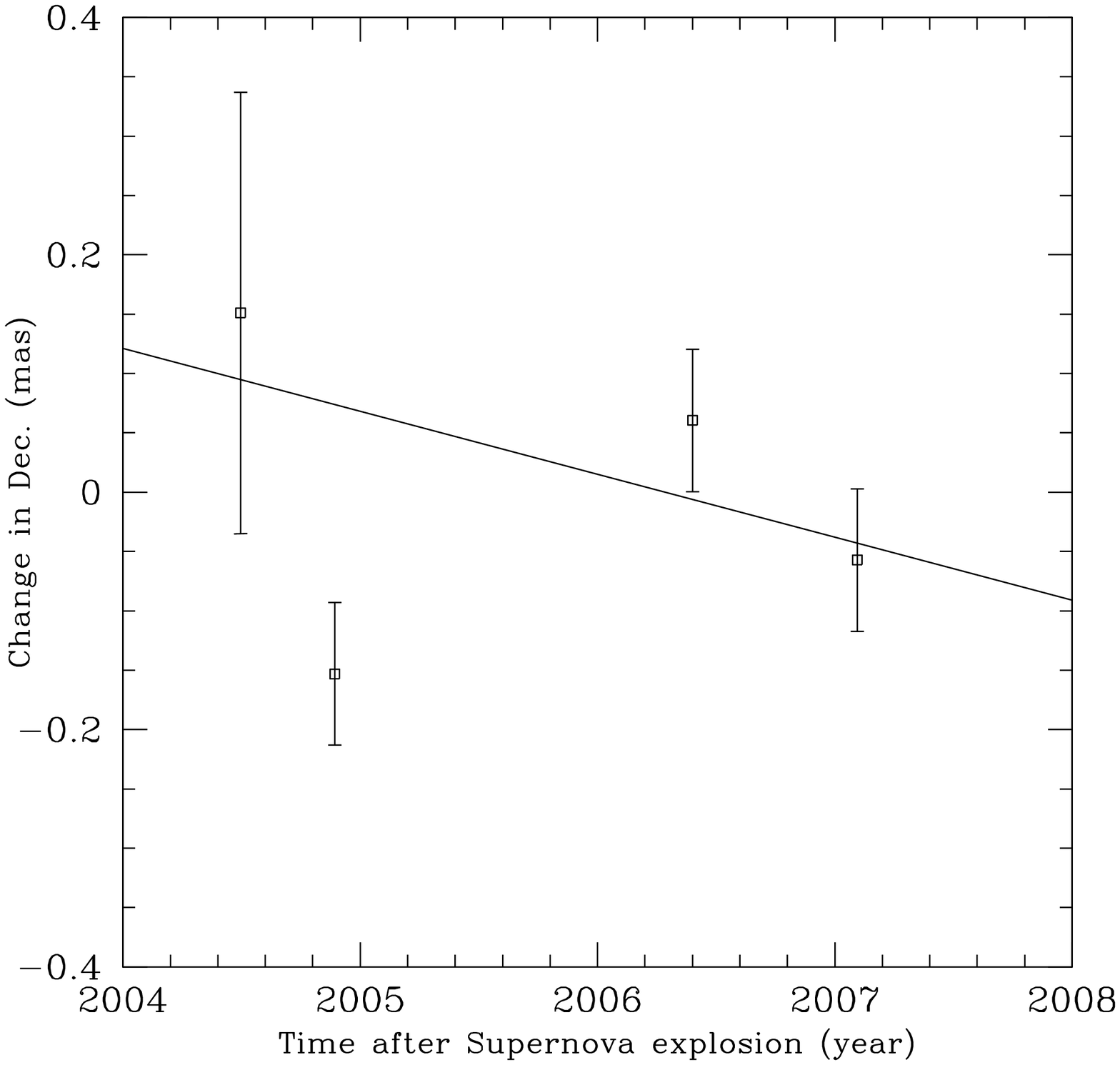}
		\figcaption{\label{fig:wlsq} Weighted least square fits for position vs. time in R.A. and Dec. This gives a proper motion for SN 2001em of (0.030 $\pm$ 0.037)~mas~year$^{-1}$ in R.A. and (-0.053 $\pm$ 0.071)~mas~year$^{-1}$ in Dec.}
	\end{figure}

		This high error is not very satisfactory and yielded unreasonably small $\chi^2$ values. As shown by \citet{2006A&A...452.1099P}, the linear approach is not reliable. Thus we had to find a better way to obtain a reasonable error for the positions. Using the discussion of astrometric accuracies of VLBA observations by \citet{2006A&A...452.1099P} and the $\chi^2$ value of the weighted least squares fit, we were able to reduce the position errors to 0.013 mas and 0.051 mas in R.A. and Dec. respectively. Using a weighted least squares fit for the motion in R.A. and Dec. separately (Fig.~\ref{fig:wlsq}), we get an error of 0.037 mas year$^{-1}$ in R.A. and 0.071 mas year$^{-1}$ in Dec. 
		
		We obtained a proper motion of (0.030 $\pm$ 0.037) mas year$^{-1}$ in R.A. and ($-$0.053 $\pm$ 0.071)~mas~year$^{-1}$ in Dec., which gives a total proper motion of 
		\begin{equation}
		 	\mu = (0.061 \pm 0.080)\,\mathrm{mas}\,\mathrm{year}^{-1}\,. 
		\end{equation}
		The proper distance to SN 2001em is 83.14 Mpc, then the proper motion gives us a projected velocity on the sky of 
		\begin{equation}
			v = (24,000 \pm 32,000)\, \mathrm{km}\,\mathrm{s}^{-1}\,.
		\end{equation}
		\cite{2007arXiv0706.3344B} report a proper motion, based on two epochs, of (0.089 $\pm$ 0.093)~mas~year$^{-1}$ or (33,000 $\pm$ 34,000)~km~s$^{-1}$, respectively, where for the cosmological parameters we use the latter becomes (35,000 $\pm$ 37,000)~km~s$^{-1}$.

	\subsection{Size Limits of the Radio Emitting Region}
		
\begin{figure}
		\plotone{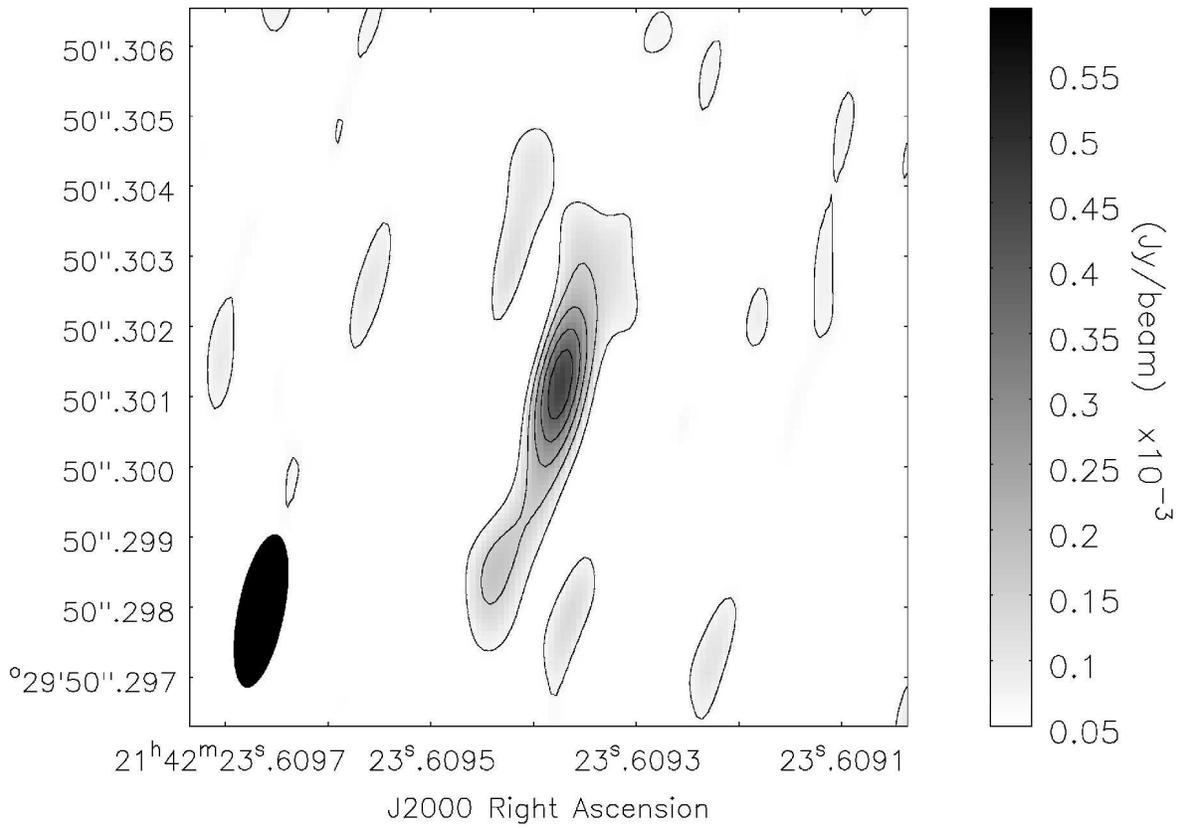}
		\figcaption{\label{fig:image} VLBI image of SN 2001em from February 4th, 2007. Peak brightness is 586 $\mu$Jy bm$^{-1}$, the rms background noise was 21 $\mu$Jy beam$^{-1}$. The contours are drawn at levels 11, 25, 39, 53, 67, 81, 95 and times 1\% of the peak brightness. The lowest contour is at 3.1$\sigma$. The grey scale flux range is 550 $\mu$Jy~beam$^{-1}$. The beamsize is shown in the lower left corner. }
	\end{figure}

		Again we used the fully calibrated and phase-referenced to J2145+1115/J2139+1423 ($u$,$v$)~plane data, to image the SN using Difmap. Fig.~\ref{fig:image} shows the VLBI image of SN 2001em on 2007, February 5. The source is not resolved. The resolution of our observation was (2.02 $\times$ 0.858) mas at $-7.17^{\circ}$. 

		A precise determination of the upper limit for the size of SN 2001em was obtained using Monte-Carlo simulation. We again used model fitting in the ($u$,$v$) plane with a circular Gaussian model. We created 100 ($u$,$v$) datasets for a range of sizes for each epoch based on flux density and corresponding image RMS noise using AIPS task UVMOD. This creates simulated datasets with similar properties to the actual data. We found the minimum $\chi^2$ for each simulated dataset and examined the Gaussian distribution of modelfit sizes depending on the simulated source diameter defined in UVMOD. This has been done for all 4 epochs and a range of simulated source sizes. The combined results of the upper limits for size (FWHM diameter of the circular Gaussian) with a confidence of 95\% are listed in Table~\ref{tab:data}. This means that 95\% of the Gaussian distribution is below the upper size limit thus giving us a 2$\sigma$ upper limit.

\begin{figure}
		\plotone{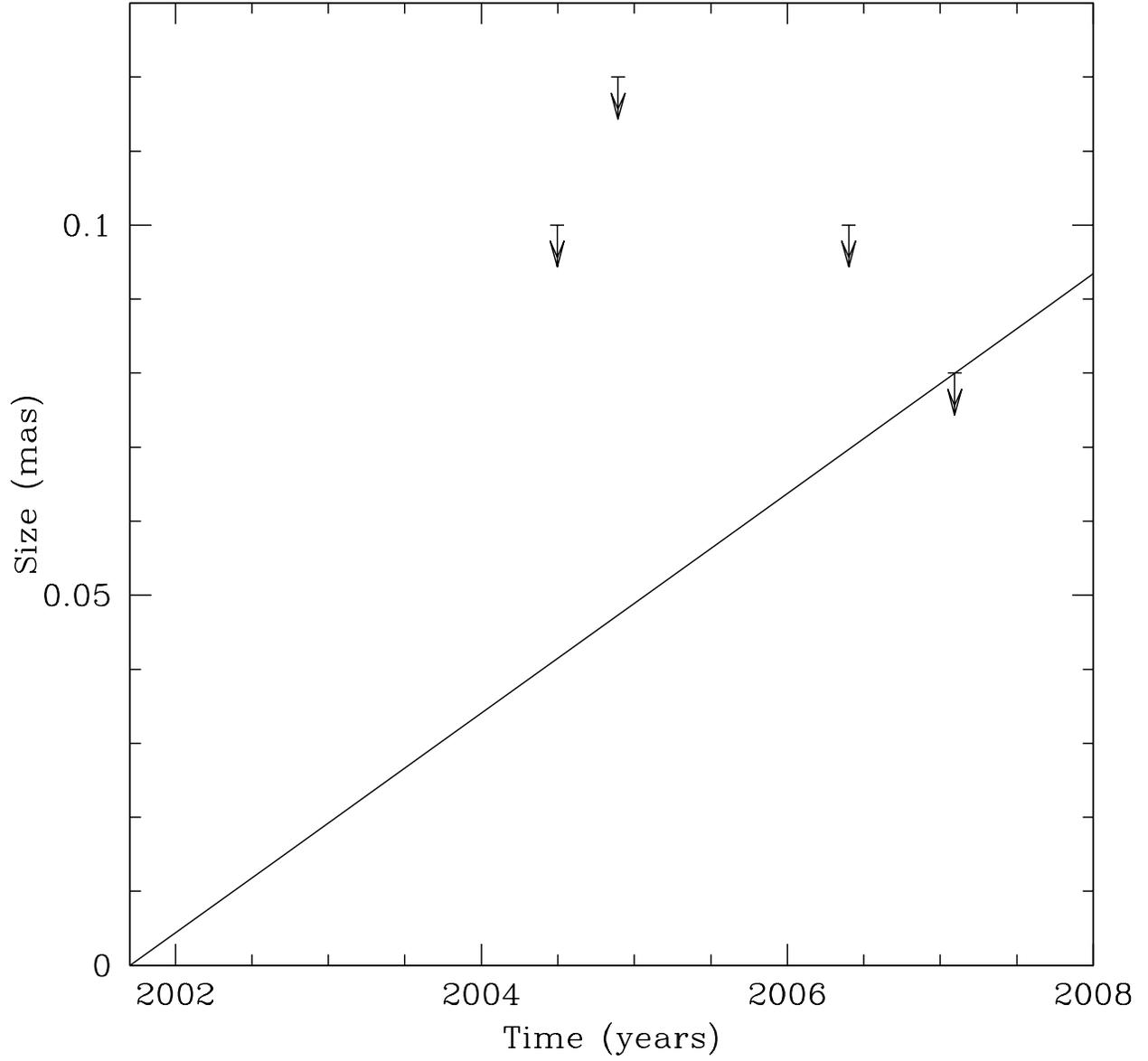}
		\figcaption{\label{fig:sizelimit} 2$\sigma$ upper limits for the radius of SN 2001em plotted over time. The linear curve represents the 2$\sigma$ upper limit for the expansion velocity of 6000 km s$^{-1}$ plotted from 2001.7041.}
	\end{figure}

		In Fig.~\ref{fig:sizelimit} we show a plot of the upper limit radii vs. time and the linear fit to obtain an upper limit for the expansion velocity. Assuming this linear expansion behavior since its discovery on 2001.704, we find a 2$\sigma$ upper limit for the expansion velocity of 0.015~mas~year$^{-1}$, which corresponds to
		\begin{equation}
			\mathrm{v}_{\mathrm{exp}} \leq 6,000\,\mathrm{km}\,\mathrm{s}^{-1}.
		\end{equation}
		\cite{2007arXiv0706.3344B} report a 2$\sigma$ upper limit for the average expansion velocity since the explosion of 25,800~km~s$^{-1}$, at 4.6 years after the SN. For that same epoch, when reanalyzing the same data we obtain a 2$\sigma$ upper limit of 8,100~km~s$^{-1}$. In contrast to \citet{2007arXiv0706.3344B} we used a circular Gaussian model for model fitting instead of an elliptical one which already reduces the size limit by a factor of about $2$. In addition, we fit directly to the visibility data and used the Monte Carlo simulation method which in the case of an unresolved source improves results further compared with image plane fitting the elliptical Gaussian using the AIPS task JMFIT.

	\subsection{Radio Lightcurve and Spectrum}
		
\begin{figure}
		\plotone{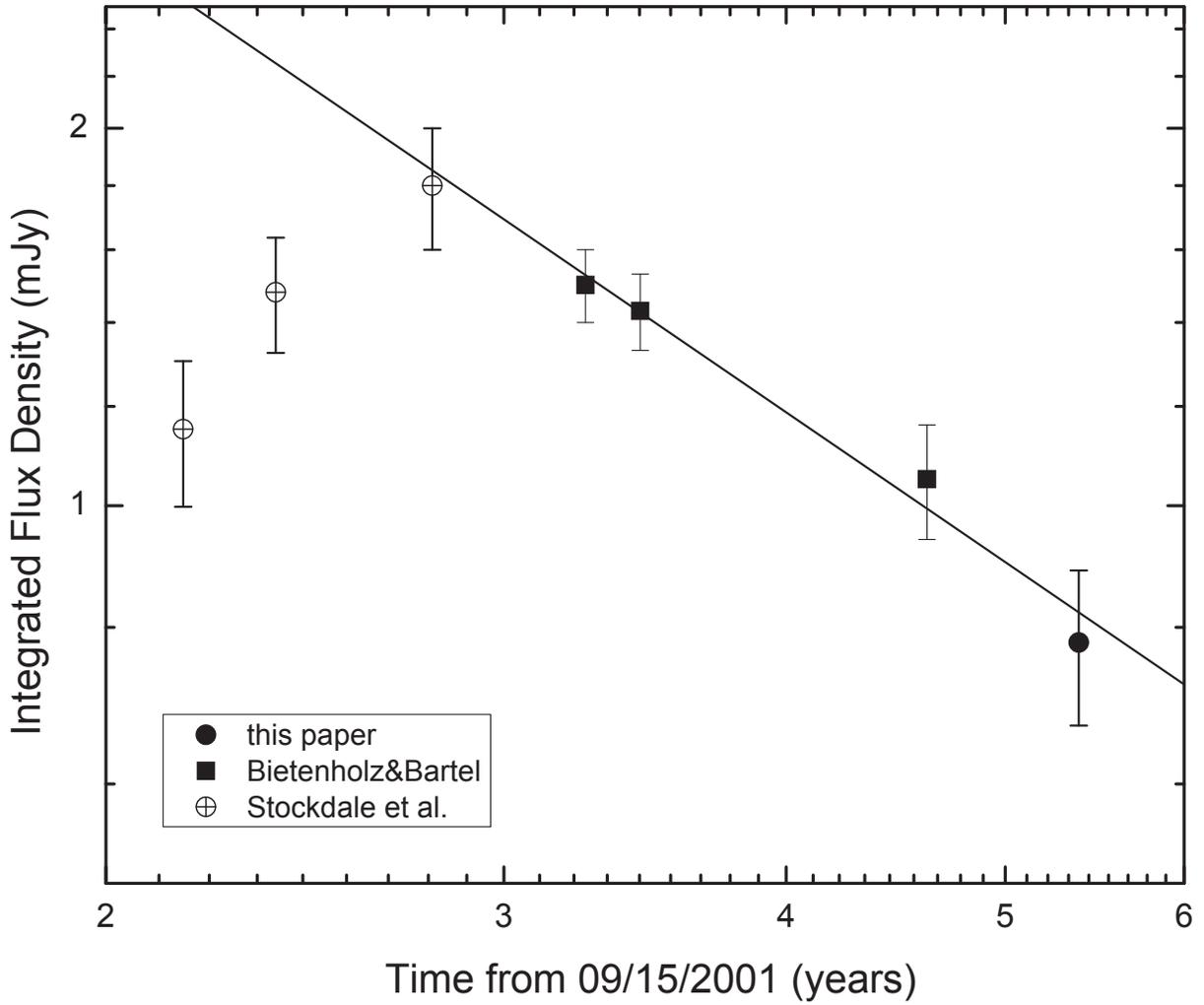}
		\figcaption{\label{fig:8ghz-lc} Plot of the 8.46 GHz lightcurve using our data from February 2007 and previously published values \citep{2007arXiv0706.3344B, 2004IAUC.8282....2S, 2005IAUC.8472....4S}. The plot shows a power law decay in the lightcurve since the peak around 2003. The slope of the power law fit is $y \propto t^{-1.23\pm0.40}$.}
	\end{figure}

		Many multifrequency observations of SN~2001em have been conducted with the VLA since 2003. We present the published values and the observations from February 2007. In Fig.~\ref{fig:8ghz-lc} we show the 8.46~GHz detections plotted over time since the supernova explosion. From the peak brightness the 8.46~GHz peak luminosity is (1.54 $\pm$ 0.10)$\times 10^{28}$ erg s$^{-1}$ Hz$^{-1}$. We see a power law decay in the flux densities since its peak around 2004 (or $\sim 2.7\;$yr since the SN) with an index of $\alpha = 1.23 \pm 0.40$, i.e. $F_\nu \propto t^{-\alpha}$.

\begin{figure}
		\plotone{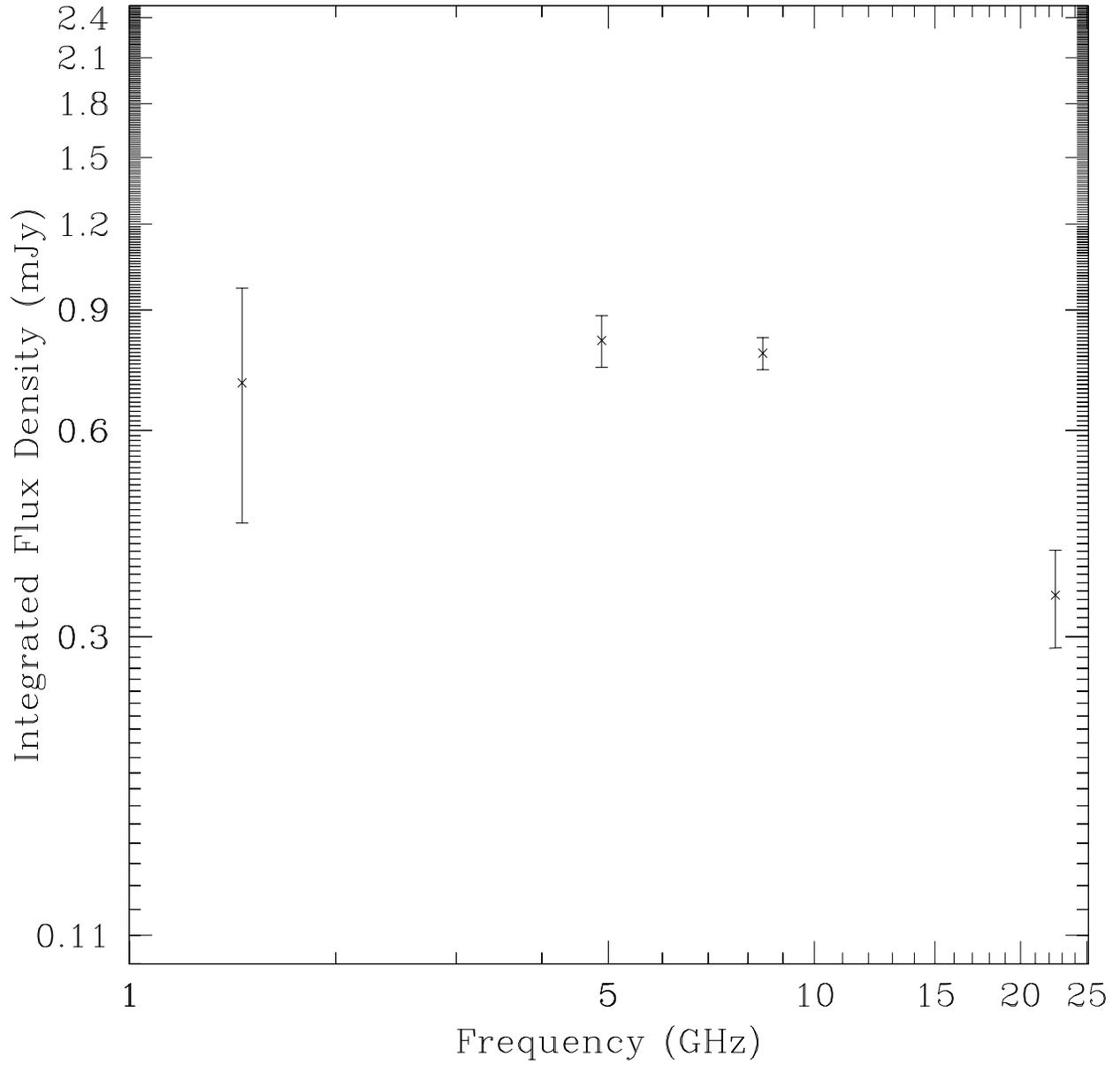}
		\figcaption{\label{fig:feb07-spec} The observed radio spectrum using the 8 GHz flux density from the phased VLA as of February 4 2007 and 1.2 GHz, 5 GHz and 22 GHz observed on February 18th, 2007.}
	\end{figure}

		Fig.~\ref{fig:feb07-spec} shows the spectrum at the time of our observations ($t = 2007.094,2007.132$) where our measurements are at 1.46~GHz, 4.9~GHz, 8.4~GHz and 22~GHz, the values are listed in Table~\ref{tab:datafeb7}. The 22~GHz point is a 5$\sigma$ detection. The spectral index ($F_\nu \propto \nu^{-\beta}$) at the cut-off between 8.4~GHz and 22~GHz is $\beta_{\mathrm{drop}} = 0.38 \pm 0.18$ and between 4.9 and 8.4~GHz it is $\beta_{\mathrm{plateau}} = 0.077 \pm 0.011$.

\section{Discussion}

		The lack of any apparent proper motion is consistent with other VLBI measurements of SNe. Some of these other VLBI monitored SNe include SNe~1986J (IIn), 1987a (II peculiar), 1993J (IIb), 2001gd (IIb), and 2004et (IIP) \citep{pereztorres02,manchester02,marcaide08,pereztorres05,martividal07}. With a fourth epoch, we have further constrained the value reported to (24,000$\pm$32,000)~km~s$^{-1}$. While such constraints on the proper motion may sometimes indicate a symmetric blast-wave centered on the supernova position, in our case the 2$\sigma$ upper limit on the average proper velocity (87,000 km s$^{-1} \approx 0.29c$) is much larger than that on the average expansion velocity (6000 km s$^{-1}$), and therefore no proper motion is expected to be detected even for a reasonably asymmetric supernova (i.e. given the limit on the expansion velocity, we would expect to detect proper motion only for an extremely asymmetric explosion, which has not yet been observed in ordinary supernovae).

		The constraint on the proper motion does, however, rule out a relativistic jet pointing away from us \citep{2004ApJ...609L...9G} as the source of the radio (and X-ray) emission.  In such a scenario, the current limit on the size of the emitting region allows only for a very compact emission region, and rules out a double sided jet where both sides are visible (even if the emission from each side is from a compact region). Moreover, the limit on the proper motion also rules out the possibility that only one side of the jet is visible, since for a compact relativistic emission region moving at an angle $\theta$ relative to our line of sight, the observed limit on the apparent velocity $\beta_{ap} = \beta\sin\theta/(1-\beta\cos\theta) < 0.29$ implies, e.g., $\theta < 1.86^\circ = 0.032\;$rad for $\beta > 0.9$. Such a small angle relative to our line of sight would not produce an optically thin rising flux $\sim 2$ years after the SN (since by that time the Lorentz factor is no more than a few and our line of sight would be well within the beaming cone of the radio emission), which was the original motivation for suggesting such an explanation.

		Comparing results for SN~2001em to SN/GRB pairs 1998bw, 2003lw and 2006aj, peak luminosities at first suggest a possible GRB event of SN~2001em, but its peak late in time does not match the pattern of SN/GRB pairs \citep{kaneko07}. In addition GRB SN show much larger expansion velocities, $>$100,000 km s$^{-1}$, than what we obtained \citep{2007ApJ...664..411P}. Moreover, our results rule out even a mildly relativistic outflow with an energy comparable to that of the SN ejecta (which is $\gtrsim 10^{51}\;$erg), which would not be capable of producing a GRB, and may potentially be much more common in core collapse SNe compared to highly relativistic outflows of similar energy \citep{2004ApJ...609L...9G}.

		Our 1$\sigma$ limiting value for the velocity of the radio-emitting region of a blastwave of $\sim\,$3,000~km~s$^{-1}$ is unusually low (see Table~\ref{tab:velocities}) and is close to the lowest measured velocity of the radio emitting region of any SN, reported for SN~1987A by \citet{manchester02} of $\sim\,$3,500~km~s$^{-1}$. \citet{martividal07} report a minimum, measured expansion velocity of (15,700 $\pm$ 2,000)~km~s$^{-1}$ for the radio emitting region of the type IIP SN~2004et. \citet{pereztorres05} reported an upper limit for the expansion velocity for the radio emitting region of the type IIb SN~2001gd of 23,000~km~s$^{-1}$.  \citet{marcaide08} report an evolving velocity for the radio-emitting region with the outer velocity of the radio emitting shell to be $\sim$~10,000~km~s$^{-1}$ and the inner velocity to be $\sim$~7,000~km~s$^{-1}$ occurring between days 1063 and 1399 since explosion. The only other type IIn SN with a measured velocity of its radio emitting region was SN~1986J, \cite{pereztorres02} report an average expansion velocity that dropped from 7,500~km~s$^{-1}$ in 1988 to 6,300~km~s$^{-1}$ in 1999.

		What is of particular interest is that SN~2001em, originally classified as s type Ic SN, has such a low average expansion velocity. \citet{soderberg03} estimate that the radio emitting region's expansion velocity if similar to the type Ic SN~2003L, should exceed 16,000~km~s$^{-1}$ and likely be as high as 30,000~km s$^{-1}$. Assuming the early optical classification of \citet{2001IAUC.7737....3F} to be valid, we might expect the blastwave of SN~2001em to have been resolved near the limit we establish had it undergone a period of rapid expansion prior to hitting a very dense CSM. Preliminary models of VLA observations suggest an average mass-loss of $\sim2.5\times 10^{-4}$~M$_\odot$~year$^{-1}$, assuming wind-driven, clumpy/filamentary CSM \citep{weiler02,2007AAS...21110521S}. \citet{weiler02} indicates that such a radio-derived mass-loss rate is slightly higher than values determined for other type IIn SNe, 1.14$\times 10^{-4}$~M$_\odot$~year$^{-1}$ for SN~1988Z and 4.28$\times 10^{-5}$~M$_\odot$~year$^{-1}$ for SN~1986J. However, depending on when this shell of material was ejected, the actual mass-loss event was likely much higher than this average value.  Due to a lack of any radio observations within the first two years following the explosion, we are unable to place any reasonable upper-bound on this mass-loss rate.

		We obtained a 2$\sigma$ upper size limit of the radio emitting region of $0.98 \times 10^{17}$ cm after 5.4 years, which is marginally consistent with the model of \citet{2006ApJ...641.1051C}, in which the ejecta collides with a dense circumstellar shell at a radius of $\sim 6\times 10^{16}$ cm after $\sim 2.6\;$yr (note that a thin emitting shell, that is more relevant for their model, gives an upper limit on the radius and average expansion velocity that is smaller by $\sim 10\%$ compared to the circular Gaussian that we have used). Our results support the picture that at a very early epoch, this supernova has evolved from a type Ib/c SN with little circumstellar interaction to a type IIn with strong circumstellar interaction. The classification as type IIn SN was in part due to the late-time observation of a narrow H$\alpha$ line \citep{2004GCN..2586....1S}. This detection indicates the existence of cool shocked circumstellar gas \citep{2006ApJ...641.1051C}. 

		The large X-ray luminosity of $L_X \approx 10^{41}\;{\rm erg\;s}^{-1}$ at $t \approx 950\;$days implies a radiated energy in the X-ray range ($0.5\,$--$\,8\;$keV) of the order of $t\,L_X(t) \sim 10^{49}\;$erg (assuming the typical time for variation in $L_X$ is $\Delta t \sim t$). This sets a strict lower limit on the energy that was dissipated and converted into internal energy by that time, of $E_{\rm dis}(t) \gtrsim 10^{49}\;$erg. However, the true value of $E_{\rm dis}(t)$ is most likely significantly higher than this lower limit, due to the combination of various inefficiencies in the conversion of the dissipated energy into radiation in the X-ray range within the dynamical time. First, the total radiated energy is probably somewhat larger than the measured value, since $\nu F_\nu$ is still rising within the observed energy range \citep[$\nu F_\nu \propto \nu^{0.9 \pm 0.35}$;][]{2004IAUC.8323....2P} and should therefore peak at higher energies. Second, not all of the dissipated energy is given to the electrons, and most of it can go into protons (or other ions; both the thermal population or cosmic rays) or magnetic fields, and thus would not be radiated away. Finally, even the energy that does go into electrons is not always radiated away efficiently on the dynamical time. Given all the above, we consider $E_{\rm dis}(t) \gtrsim 10^{51}\;$erg to be a more realistic estimate of the energy that has been dissipated by time $t$ (although a somewhat lower value might still be possible under some circumstances). This would suggest a deceleration time of $t_{\rm dec} \lesssim 10^3\;$days, i.e. that by that time the ejecta had interacted with a surrounding mass comparable to its own mass \citep{2004ApJ...609L...9G,2007RMxAC..30...41C}.

		Moreover, the radio light curve also suggests a deceleration time of $t_{\rm dec} \sim 10^3\;$days, while our limits on the source size imply that the typical initial velocity $v_0$ of the mass that has decelerated by $t_{\rm dec}$ is $\lesssim 8,000\;{\rm km\;s^{-1}}$. For an external density $\rho = Ar^{-2}$ this implies $A_* = A/(5\times 10^{11}\;{\rm g\;cm^{-1}}) \gtrsim 7.3\times 10^3(E/10^{51}\,{\rm erg})(t_{\rm dec}/10^3\,{\rm days})^{-1}(v_0/8\times 10^3\;{\rm km\;s^{-1}})^{-3}$ that corresponds to a mass-loss rate of $\dot{M} \gtrsim 0.073(v_w/10^3\,{\rm km\;s^{-1}})(E/10^{51}\,{\rm erg})(t_{\rm dec}/10^3\,{\rm days})^{-1}(v_0/8\times 10^3\;{\rm km\;s^{-1}})^{-3}\;M_\odot\;{\rm yr^{-1}}$ where $v_w$ is the stellar wind velocity of the progenitor star, assuming a quasi steady wind during the relevant time scale. Such a high mass loss rate may suggest an episodic mass ejection event, rather than a steady wind, that would result in a dense CSM shell.  The required mass of such a CSM shell is at least comparable to that of the SN ejecta that was decelerated until $t_{\rm dec}$, $M_{\rm shell} \gtrsim M_0 = 2E/v_0^2 = 1.57(E/10^{51}\,{\rm erg})(v_0/8\times 10^3\;{\rm km\;s^{-1}})^{-2}\;M_\odot$. Now, let us consider the opacity of this part of the original SN ejecta, of mass $M_0$ and initial velocity $v_0$, at earlier times. Assuming one free electron per proton mass, its Thompson optical depth at $t < t_{\rm dec}$ would be $\tau_T = \sigma_T E/(2\pi m_p v_0^4 t^2) = 23 (E/10^{51}\,{\rm erg})(v_0/8\times 10^3\;{\rm km\;s^{-1}})^{-4}(t/30\,{\rm days})^{-2}$, while additional opacity from other parts of the ejecta could only add to the total optical depth. This opacity is somewhat large for the time near the peak of the supernova optical light curve ($t \sim 30\,$days), and may become a more severe problem if $v_0$ turns out to be lower than the current upper limit on it.

\section{Summary \& future observations}

		It is clear that SN~2001em exhibits radio properties consistent with a type II SN with a significant CSM/blastwave interaction.  Future VLBI monitoring is recommended. As the blastwave expands we will be able to further constrain or eventually determine the average velocity of the blastwave. SN~2001em has faded considerably at 8~GHz over the last few years but is still readily detectable with a estimated flux density at the end of 2008 of 0.55 mJy. With HSA recording rates expected to increase to 4096 Mbps by 2011, the thermal noise level of the HSA at 8 GHz should decrease to $\sim$1.40 $\mu$Jy beam$^{-1}$ at an integration time of 8 hours, thus keeping pace with the fading of SN 2001em which is expected to follow a power law decay. At the end of 2008 an HSA observation (512 Mbit s$^{-1}$) would have a signal to noise ratio (SNR) of 139 and by 2011 with 4096 Mbit s$^{-1}$ bandwidth a SNR of 288 based on 1$\sigma$ using natural weighting.

		\acknowledgements
        The National Radio Astronomy Observatory is a facility of the National Science Foundation operated under cooperative agreement by Associated Universities, Inc. J.G. gratefully acknowledges a Royal Society Wolfson Research Merit Award.  E. R. acknowledges support from the DOE Program for Scientific Discovery through Advanced Computing (SciDAC; DE-FC02-01ER41176). C. J. S. is a Cottrell Scholar of the Research Corporation. We thank an anonymous referee for constructive comments.

	{}

\pagebreak

	\begin{deluxetable}{lllllll}
		\tabletypesize{\scriptsize}
		\tablecaption{\label{tab:observation} Observational Summary}
		\tablehead{
			\colhead{Date} & \colhead{Time after SN} & \colhead{Frequency} & \colhead{Integration} & \colhead{Bandwidth} & \colhead{Polarization} & \colhead{Instrument} \\
			 & & & \colhead{Time} \\
			 & \colhead{(days)} & \colhead{(GHz)} & \colhead{(minutes)} & \colhead{(MHz)}
		}
		\startdata
			2004 Jul 01 ...... & 1021   & 8.410  & 178  & 32 & 2 & VLBA \\
			2004 Nov 22 ...... & 1165   & 8.410  & 400  & 32 & 2  & VLBA+AR+EB+GBT+Y27\\
			2006 May 27 ...... & 1686   & 8.406  & 423  & 32 & 2 & VLBA+AR+EB+GBT+Y27\\
			2007 Feb 04 ...... & 1969   & 8.410  & 300 & 32 & 2 & VLBA+AR+EB+GBT+Y27\\
			2007 Feb 18 ...... & 1983   & 1.465  & 12 & 50 & 2 & VLA\\
					   &        & 4.885  &  9 & 50 & 2 & VLA\\
					   &        & 22.485 & 20 & 50 & 2 & VLA
		\enddata
		\tablenotetext{\,}{Notes.- AR = 305 m Arecibo telescope. EB = 100 m Effelsberg telescope. GBT = 105 m GBT. Y27 = phased VLA.}
	\end{deluxetable}

	\begin{deluxetable}{llllll}
		\tabletypesize{\scriptsize}
		\tablecaption{\label{tab:data} Results for the 2$\sigma$ upper size limit of the FWHM diameter of a circular Gaussian point source, obtained from Monte Carlo simulation. The position offsets are given using the reference position 21h42m23.619s R.A. and +12d29m50.299s Dec.}
		\tablehead{
			\colhead{Epoch} & \colhead{Flux Density} & \colhead{Offset R.A.} & \colhead{Offset Dec.} & \colhead{RMS} & \colhead{Sizelimit} \\
			\colhead{(years)} & \colhead{(mJy)} & \colhead{(mas)} & \colhead{(mas)} & \colhead{($\mu$Jy/beam)} & \colhead{(mas)}
		}
		\startdata
			2004.496 & 1.41 & -0.089 & 0.15 & 59 & 0.19 \\
			2004.893 & 0.60 & 0.028 & -0.15 & 36 & 0.23 \\
			2006.401 & 0.52 & 0.083 & 0.061 & 21 & 0.19 \\
			2007.094 & 0.59 &  -0.019 & -0.057 & 21 & 0.16 \\
		\enddata
	\end{deluxetable}

	\begin{deluxetable}{llllll}
		\tabletypesize{\scriptsize}
		\tablecaption{\label{tab:datafeb7} Results for flux densities and corresponding noise of SN 2001em on February 18, 2007.}
		\tablehead{
			\colhead{Frequency} & \colhead{Flux Density} & \colhead{Noise} \\
			\colhead{(GHz)} & \colhead{(mJy)} & \colhead{(mJy)}
		}
		\startdata
			1.465  & 0.704 & 0.264 \\
			4.885  & 0.812 & 0.070 \\
			22.485 & 0.345 & 0.056 \\
		\enddata
	\end{deluxetable}

	\begin{deluxetable}{llll}
        \tabletypesize{\scriptsize}
        \tablecaption{\label{tab:velocities} Expansion Velocities for Type II Supernovae.}
        \tablehead{
            \colhead{Supernova} & \colhead{Type} & \colhead{Velocity} & \colhead{Reference} \\

             &  & \colhead{($10 ^3$ km s$^{-1}$)} &
        }
        \startdata
            1987A                  & IIpec  & $\sim 3.5$      & \citet{manchester02} \\
            2004et                             & IIP    & $> 15.7\pm 2.0$ & \citet{martividal07}  \\
                        2001gd                 & IIb    & $23.0$          & \citet{pereztorres05} \\
                        1993J (outer shell)    & IIb    & $\sim 10.0$      & \citet{marcaide08} \\
                        1993J (inner shell)    & IIb    & $\sim 7.0$       & \citet{marcaide08} \\
                        1986J (measured 1988)  & IIn    & $7.5$           & \citet{pereztorres02} \\
                        1986J (measured 1999)  & IIn    & $6.3$           & \citet{pereztorres02} \\
            				2001em (measured 2006) & IIn    & $< 8.1$         & this manuscript \\
                        2001em (measured 2007) & IIn    & $< 6.0$         & this manuscript

         \enddata
    \end{deluxetable}

	
\begin{thebibliography}{}

		\bibitem[Bietenholz \& Bartel(2005)]{2005ApJ...625L..99B} Bietenholz, M.~F., \& Bartel, N.\ 2005, \apjl, 625, L99 

		\bibitem[Bietenholz \& Bartel(2007)]{2007arXiv0706.3344B} Bietenholz, M.~F., \& Bartel, N.\ 2007, ArXiv e-prints, 706, arXiv:0706.3344

		\bibitem[Chevalier(2007)]{2007RMxAC..30...41C} Chevalier, R.~A.\ 2007, Revista Mexicana de Astronomia y Astrofisica Conference Series, 30, 41  

		\bibitem[Chugai \& Chevalier(2006)]{2006ApJ...641.1051C} Chugai, N.~N., \& Chevalier, R.~A.\ 2006, \apj, 641, 1051 

		\bibitem[Feissel-Vernier(2003)]{2003A&A...403..105F} Feissel-Vernier, M.\ 2003, \aap, 403, 105 
		
		\bibitem[Filippenko \& Chornock(2001)]{2001IAUC.7737....3F} Filippenko, A.~V., \& Chornock, R.\ 2001, \iaucirc, 7737, 3 

		\bibitem[Granot et al.(2002)]{Granot02} Granot, J., Panaitescu, A., Kumar, P., \& Woosley, S.~E. 2002, ApJ, 570, L61

		\bibitem[Granot \& Loeb(2003)]{GL03} Granot, J., \& Loeb, A. 2003, ApJ, 593, L81 

		\bibitem[Granot \& Ramirez-Ruiz(2004)]{2004ApJ...609L...9G} Granot, J., \& Ramirez-Ruiz, E.\ 2004, \apjl, 609, L9 

		\bibitem[Hinshaw et al.(2008)]{2008arXiv0803.0732H} Hinshaw, G., et al.\ 2008, ArXiv e-prints, 803, arXiv:0803.0732 

		\bibitem[Kaneko et al.(2007)]{kaneko07} Kaneko, Y., et al.\ 2007, \apj, 654, 385 

		\bibitem[Marcaide et al.(2008)]{marcaide08}  Marcaide, J.~M., Mart{\'{\i}}-Vidal, I., Alberdi, A., P{\'e}rez-Torres, M.~A., Ros, E., Diamond, P. J., Guirado, J.~C., Lara, L., Shapiro, I.~I., Stockdale, C.~J., Weiler, K.~W., Mantovani, F., Preston, R. A., Schlizzi, R. T., Sramek, R. A., Trigilio, C., Van Dyk, S. D., \& Whitney, A. R.\ 2007, in prep.

		\bibitem[Mart{\'{\i}}-Vidal et al.(2007)]{martividal07} Mart{\'{\i}}-Vidal, I., Marcaide, J.~M., Alberdi, A., Guirado, J.~C., Lara, L.,P{\'e}rez-Torres, M.~A., Ros, E., Argo, M.~K., Beswick, R.~J., Muxlow, T.~W.~B., Pedlar, A., Shapiro, I.~I., Stockdale, C.~J., Sramek, R.~A., Weiler, K.~W., \& and Vinko, J.\ 2007, \aap, 470, 1071
		
		\bibitem[Manchester et al.(2002)]{manchester02} Manchester, R. N., Gaensler, B. M., Wheaton, V. C., Stavely-Smith, L., Tzioumis, A. K., Bizunok, N. S., Kesteven, M. J., \& Reynolds, J. E.\ 2002, Publications of the Astronomical Society of Australia, 19, 207

		\bibitem[Paczy\'nski(2001)]{Pacz01} Paczy\'nski, B. 2001, Acta Astron., 51, 1
		
		\bibitem[Papenkova \& Li (2001)]{2001IAUC.7722....1P} Papenkova, M., Li, W.~D., Wray, J., Chleborad, C.~W., \& Schwartz, M.\ 2001, \iaucirc, 7722, 1
		
		\bibitem[Paragi et al.(2005)]{2005MmSAI..76..570P} Paragi, Z., Garrett, M.~A., Paczy{\'n}ski, B., Kouveliotou, C., Szomoru, A., Reynolds, C., Parsley, S.~M., \& Ghosh, T.\ 2005, Memorie della Societa Astronomica Italiana, 76, 570
		
		\bibitem[P{\'e}rez-Torres et al.(2002)]{pereztorres02} P{\'e}rez-Torres, M. A., Alberdi, A., Marcaide, J. M., Guirado, J. C., Lara, L., Mantovani, F., Ros, E., Weiler, \& K. W.\ 2002, \mnras, 335, L23

		\bibitem[P{\'e}rez-Torres et al.(2005)]{pereztorres05} P{\'e}rez-Torres, M. A., Alberdi, A., Marcaide, J. M., Guirado, J. C., Lara, L., Mantovani, F., Ros, E., Weiler, K. W., \& Stockdale, C. J.\ 2005, \mnras, 360, 1055

		\bibitem[Pihlstr{\"o}m et al.(2007)]{2007ApJ...664..411P} Pihlstr{\"o}m, Y.~M., Taylor, G.~B., Granot, J., \& Doeleman, S.\ 2007, \apj, 664, 411

		\bibitem[Piner et al.(2007)]{2007AJ....133.2357P} Piner, B.~G., Mahmud, M., Fey, A.~L., \& Gospodinova, K.\ 2007, \aj, 133, 2357

		\bibitem[Pooley \& Lewin(2004)]{2004IAUC.8323....2P} Pooley, D., \& Lewin, W.~H.~G.\ 2004, \iaucirc, 8323, 2 		

		\bibitem[Pradel et al.(2006)]{2006A&A...452.1099P} Pradel, N., Charlot, P., \& Lestrade, J.-F.\ 2006, \aap, 452, 1099
		
		\bibitem[Savolainen et al.(2006)]{2006ApJ...647..172S} Savolainen, T., Wiik, K., Valtaoja, E., Kadler, M., Ros, E., Tornikoski, M., Aller, M.~F., \& Aller, H.~D.\ 2006, \apj, 647, 172
		
		\bibitem[Soderberg et al.(2003)]{soderberg03} Soderberg, A. M., Kulkarni, S. R., Berger, E., Chevalier, R. A., Frail, D. A., Fox, D. B., \& Walker, R. C.\ 2003, \apj, 621, 908

		\bibitem[Soderberg et al.(2004)]{2004GCN..2586....1S} Soderberg, A.~M., Gal-Yam, A., \& Kulkarni, S.~R.\ 2004, GRB Coordinates Network, 2586, 1

		\bibitem[Stockdale et al.(2004)]{2004IAUC.8282....2S} Stockdale, C.~J., Van Dyk, S.~D., Sramek, R.~A., Weiler, K.~W., Panagia, N., Rupen, M.~P., \& Paczynski, B.\ 2004, \iaucirc, 8282, 2

		\bibitem[Stockdale et al.(2004)]{2004AAS...205.7107S} Stockdale, C.~J., Van Dyk, S.~D., Weiler, K.~W., Sramek, R.~A., Panagia, N., Rupen, M.~P., Paczynski, B., \& Weiler, K.~W.\ 2004, Bulletin of the American Astronomical Society, 36, 1464

		\bibitem[Stockdale et al.(2005)]{2005IAUC.8472....4S} Stockdale, C.~J., et al.\ 2005, \iaucirc, 8472, 4

		\bibitem[Stockdale et al.(2007)]{2007AAS...21110521S} Stockdale, C., Kelley, M.~T., Weiler, K.~W., Panagia, N., \& Sramek, R.~A.\ 2007, American Astronomical Society Meeting Abstracts, 211, \#105.21

		\bibitem[Stockdale et al.(2008)]{stockdale08} Stockdale, C. J., Kelley, M. T., Weiler, K. W., Panagia, N., Sramek, R. A., Marcaide, J. M., Williams, C. L. M., \& Van Dyk, S. D.\ 2008, American Institute of Physics Conference Series, 937, 264

		\bibitem[Weiler et al.(2002)]{weiler02} Weiler, Kurt W. Panagia, Nino Montes, Marcos J., \& Sramek, Richard A.\ 2002, \araa, 40, 387

		
	\end{thebibliography}
\end{document}